# CT density estimation and beam hardening correction for single material objects using reference scans - a simulation and experimental study

Nikhil Deshmukh[a]

[a]Pacific Northwest National Laboratory, Richland, Washington

*E-mail: nikhil.deshmukh@pnnl.gov*

# CT density estimation and beam hardening correction for single material objects using reference scans - a simulation and experimental study

*Abstract: We describe the results from a simulation and experimental study of density estimation and beam hardening correction for Bremsstrahlung source computed tomography scan of objects made of a single material. The correction is done using transmission scans of a number of sheets of known thickness and density and made of the same material as the object under test.*

## Introduction

Industrial computed tomography (CT) scanners using a Bremsstrahlung x-ray source have certain drawbacks when using the CT reconstructions to obtain a quantitative density profile in the object under test (OUT). Although one gets a relative density profile within the OUT, the polychromatic nature of the x-ray source means that unless the spectrum is well known [1], one cannot convert the transmission measurements into an absolute density number. In addition, CT scans with Bremsstrahlung sources suffer from a well-known artifact known as beam hardening, where the interior of the object appears less dense than the outside of the object. This artifact, which occurs due to preferential absorption of low-energy x-rays and the dependence of the attenuation coefficient of a material on x-ray energy, is also known as the cupping artifact.

Correction of CT projections using reference objects like a step wedge is a well-known technique. We describe the results of a simulation study and experiment in which transmission scans of sheet references (SR, sheets made of well-characterized reference material) are used to get both bulk density and density variation within the OUT, along with beam hardening correction.

## Assumptions

The method assumes that the OUT is made of a single material (element or compound), has unknown bulk density with unknown internal density variation that we want to measure, and that SRs of the same material with precisely known thicknesses and density (or densities) are available. SR density does not have to be related to OUT density in any way. The SR and OUT must be scanned with the same detector and Bremsstrahlung spectrum, i.e. the same source (anode material, filter) with the same end point energy, source to object distance, object to detector geometry, etc. The method does not correct for scatter, other than using a scanning geometry which would minimize the effect of scatter. Although the restrictions may seem quite severe, for single material objects in an industrial setting there are situations where these requirements can be met.

## Procedure

The following describes the procedure for using measurements of SRs to make beam hardening corrected measurements of the OUT.

1. Make transmission measurements of SRs to obtain transmission fraction (commonly known as $I/I_0$ in CT literature) for a range of thicknesses. The range of density-thickness products

(also referred to as areal density because the units are g/cm$^2$) in the SRs should cover the full range of density-thickness products present in the projections of the OUT. The SR measurements should be made with an imaging geometry which reduces scatter as much as possible and preferably with the same geometry as the geometry used for the OUT.
2. Create a lookup table between the SR transmission fractions and SR density-thickness product. The transmission fractions can be averaged over a large region of the SRs if the density is not uniform. Fig.1 shows the plot of such a lookup table.

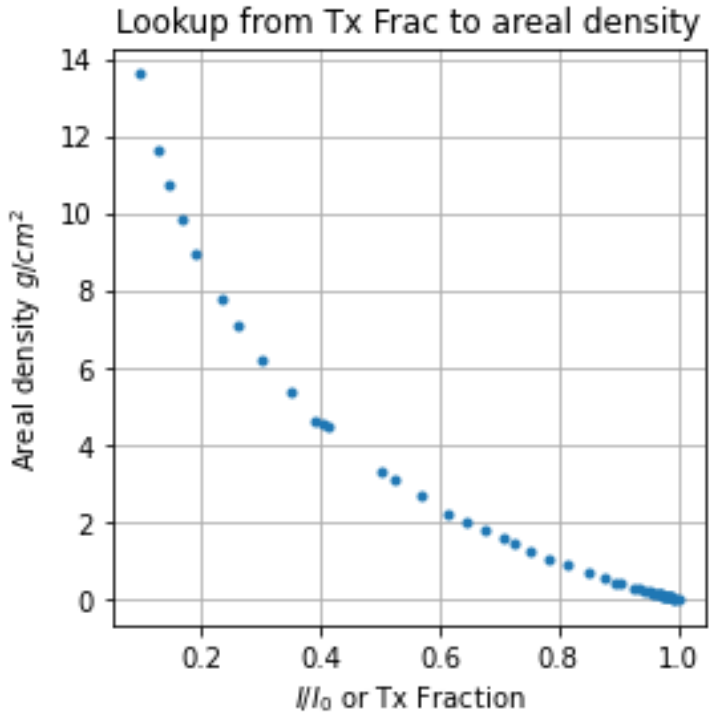

Fig 1. Plot shows the lookup table to convert the OUT projections from transmission fraction (I/I$_0$) to areal density (density-thickness product) of reference material (graphite in this case).

3. Make CT projections of OUT. Projections are the transmission fraction (I/I$_0$).
4. Use the SR lookup table to convert the OUT projections from I/I$_0$ (transmission fraction) space to areal-density space, pixel-by-pixel.

*Explanation:*
The motivation for using a lookup table can be seen from the equation describing polychromatic transmission of a ray with spatial and energy discretization:

$$I = I_{in}^1 \, e^{-MAC^1(\rho_1 l + \rho_2 l + \cdots + \rho_m l)} + I_{in}^2 \, e^{-MAC^2(\rho_1 l + \rho_2 l + \cdots + \rho_m l)} + \cdots + I_{in}^2 \, e^{-MAC^n(\rho_1 l + \rho_2 l + \cdots + \rho_m l)}$$

$I_{in}$ – The input beam intensity at each energy. (commonly known as $I_0$ when there is no risk of confusion with voxel subscripts)

$l$ – length of voxel

$\rho_x$ – density for voxel x

$MAC^y$ – mass attenuation coefficient for energy bin y

m – total number of voxels

n – total number of energy bins

Traditionally we take the log of the transmission fraction before reconstructing since in the monochromatic case, taking the log of both sides of the equation removes the exponential and linearizes the right side with respect to path length. Since this is a poor approximation in the polychromatic case, using the reference data does a better job of linearizing the projections.

To further justify this process, we can compare the line integral view of the standard CT integral with a new integral that captures the effect of the lookup table, including how the units/dimensions compare between the two.

**Standard CT line integral is defined as:**

$$\int_{Line} \mu \cdot dL$$

where µ(x,y,z) is a 3D scalar field of the linear attenuation coefficient (LAC) and has units of inverse length, and x,y,z are the spatial coordinates of the object space. When discretized, this is the voxel space.

The line(s) in the line integral are the lines or rays in object/voxel space starting from the source, intersecting the object, and going to the detector.

The Radon transform using this line integral gives us projections of ($\mu_1 * l + \mu_2 * l$ ...) along each ray when discretized, and they are dimensionless:

$$\mathcal{R}(\mu, Line) = \int_{Line} \mu \cdot dL$$

**The new line integral, called density-integral hereafter, is defined as:**

$$\int_{Line} \rho \cdot dL$$

Where $\rho$ (x,y,z) is density of the material.

The Radon transform using the density-integral gives us projections of ($\rho_1 * l + \rho_2 * l$ ...) or areal density with dimensions $ML^{-2}$.

$$\mathbb{R}(m, \text{Line}) = \int_{Line} \rho \cdot dL$$

This fits in with step 1 where we convert the OUT projections from transmission fraction to g/cm² of reference material.

**Side by side comparison of the two inverse Radon transforms and the associated units**

| Standard CT inverse Radon transform | Density-integral inverse Radon transform |
|---|---|
| $\int_{Line} \mu \cdot dL$ | $\int_{Line} \rho \cdot dL$ |
| Input projections are line integrals of µ.dL, **dimensions**: dimensionless | Input projections are line integrals of $\rho$.dL **dimensions**: ML$^{-2}$ |
| Output µ(x,y,z), a 3D scalar field **dimensions**: L$^{-1}$ | Output $\rho$(x,y,z), a 3D scalar field. **dimensions**: ML$^{-3}$ |

Inverse Radon transform and similar reconstruction algorithms can be applied to any quantity that can be integrated along a line. In standard CT, the quantity ($\mu_1$*l + $\mu_2$*l + $\mu_3$*l ...) is summed along lines. In the modified technique, the density-thickness product of SR encountered by the ray/beam are summed together ($\rho_1$*l + $\rho_2$*l + $\rho_3$*l ...).

The input/output units are different, but in both cases the inverse radon transform multiplies the input units by units of inverse length. Other than that, the computations are similar and standard CT reconstruction software can be used to do this inverse Radon transform.

5. Do a reconstruction of the transformed projections from step 3 using any CT reconstruction software. This gives us the density of reference material in each voxel in whatever units the lookup table output was created (if you use a CT software like ASTRA which outputs quantities in units of voxel length, you will have to divide by voxel length).

# Results from MCNP simulations

We use MCNP6 [2] to simulate the projection(s) of a cylindrical object made of carbon with three concentric regions with different densities. The source is programmed to emit a polychromatic spectrum, as shown in Fig 2a. The three concentric regions have densities of 1.66 g/cc, 1.76 g/cc, and 1.86 g/cc from innermost to outermost ring (densities are chosen to be similar to the densities in the experiment described in the next section). The inner cylinder has a radius of 1 cm. The middle and outer annuli have wall thicknesses of 1 cm. CardSharp [3] was used to generate the MCNP input deck.

The MCNP simulation of the projection(s) was run with an FIR5 tally, 300x300 pixels, with a 130 kVp spectrum (Fig 2a). Projections with and without scatter were simulated.

Transmission fractions (Fig 2b) using the chosen spectrum were obtained for the SRs using MCNP simulations for 15 thicknesses of carbon ranging from 0.5 cm to 7.5 cm and a density of 1.78 g/cc.

The spectrum is only used to generate the projections of the OUT and the SR transmission measurements. The spectrum is not used in the next phase where the beam hardening correction and density estimation is performed.

Once the OUT projections and SR measurements are in hand, we do the computations as described in the procedure above. Reconstructions are done using the ASTRA [1] toolbox. Fig 3 shows slices from the reconstruction. Fig 4 shows line plots from the middle of the object with and without beam hardening correction (BHC). BHC is carried out with and without scatter in the simulations.

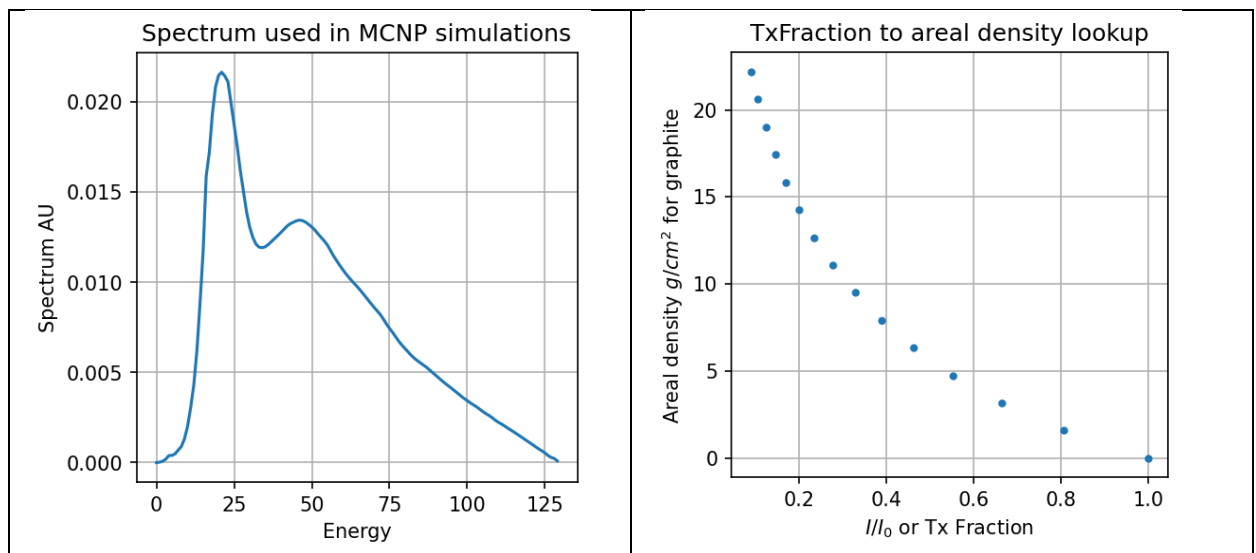

Fig 2. Left: Spectrum used in the MCNP model. Right: Lookup from transmission fraction to length of reference material (carbon at 1.78 g/cc).

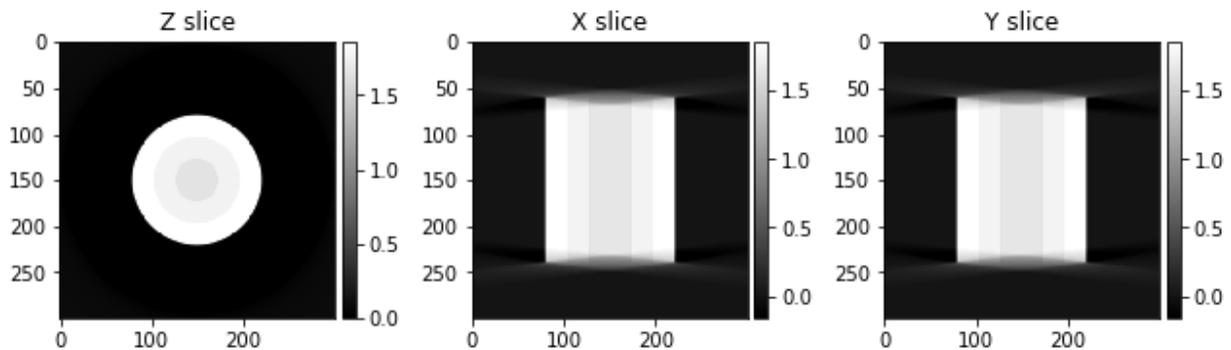

Fig 3. Slices through reconstructed object: Z (left), X (middle), and Y (right).

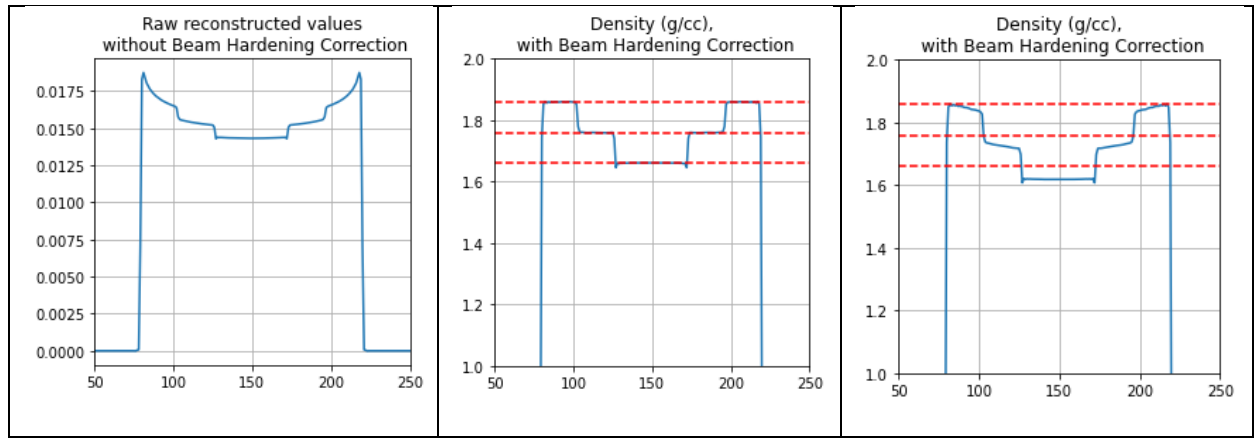

Fig 4. (Left) Line plot through the middle of the object reconstructed without beam hardening correction (raw CT reconstruction units) from OUT projections with scatter. (Middle) Line plot through the middle of the reconstruction with beam hardening correction and conversion to density (units: g/cc). The red lines show the expected densities (1.66, 1.76, and 1.86 g/cc, inside to outside). Both the OUT and SR scans are without scatter. The cupping artifact is completely corrected, and the mean densities are correctly recovered. (Right) Same as b, but both OUT and SR have scatter included. The beam hardening correction is not perfect when scatter is included (measured densities of 1.62, 1.73, and 1.85 g/cc, which correspond to 2.4%, 1.7%, and 0.53% underestimation from inside to outside).

As can be seen, the reconstructions are corrected for beam hardening, but more importantly, the estimated density values are fairly close to the real values. When scatter is included, the method underestimates the densities by <3%. This indicates the importance of reducing scatter in the setup.

Simulations were conducted for other materials with similar results.

## Results from experiment

In addition to the simulations, an experiment was also performed using sheets and cylinders of graphite as the SRs and OUT, respectively.

Four cylinders made of graphite with 1 cm diameter and different densities were scanned using a Heliscan CT scanner at 160 kVp (Fig 5).

Forty-six sheets of reference graphite material with thicknesses ranging from 1.5 cm to 7.6 cm were scanned at 160 kVp to generate the conversion table in Fig 1. The density of the reference material was declared by the manufacturer to be 1.78 g/cc.

We apply the corrections and do the reconstruction as described in the procedure above. Fig 6 shows a slice from the middle of the reconstruction.

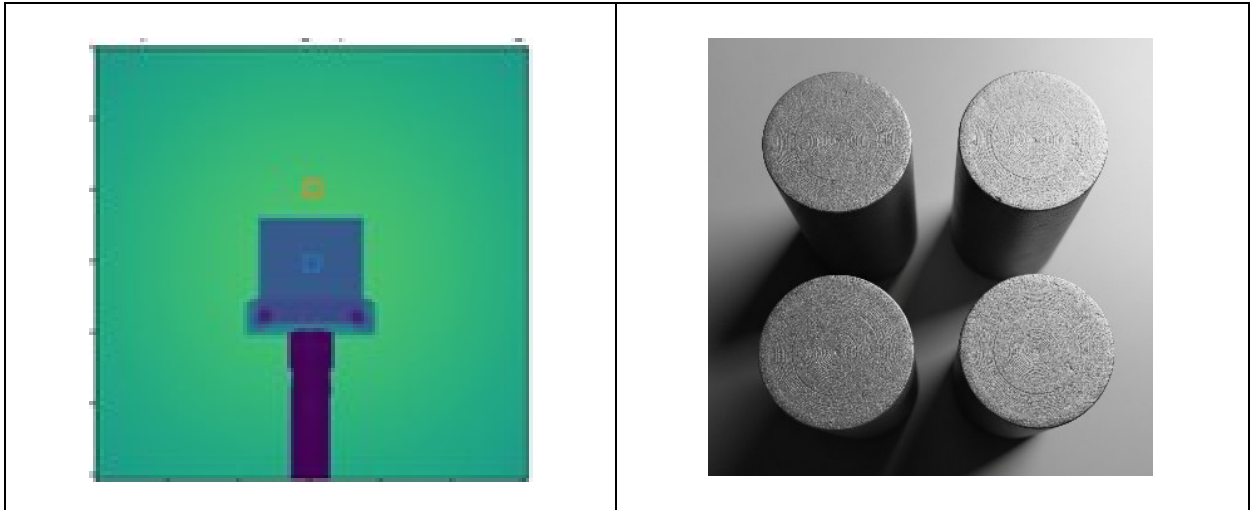

Fig 5. Left: Scan of graphite sheet in a 3D-printed holder attached to the CT sample stage. Small rectangles show sampling regions for the object and clear field. Right: Picture of graphite cylinders for illustration only (picture of scanned cylinders is not available).

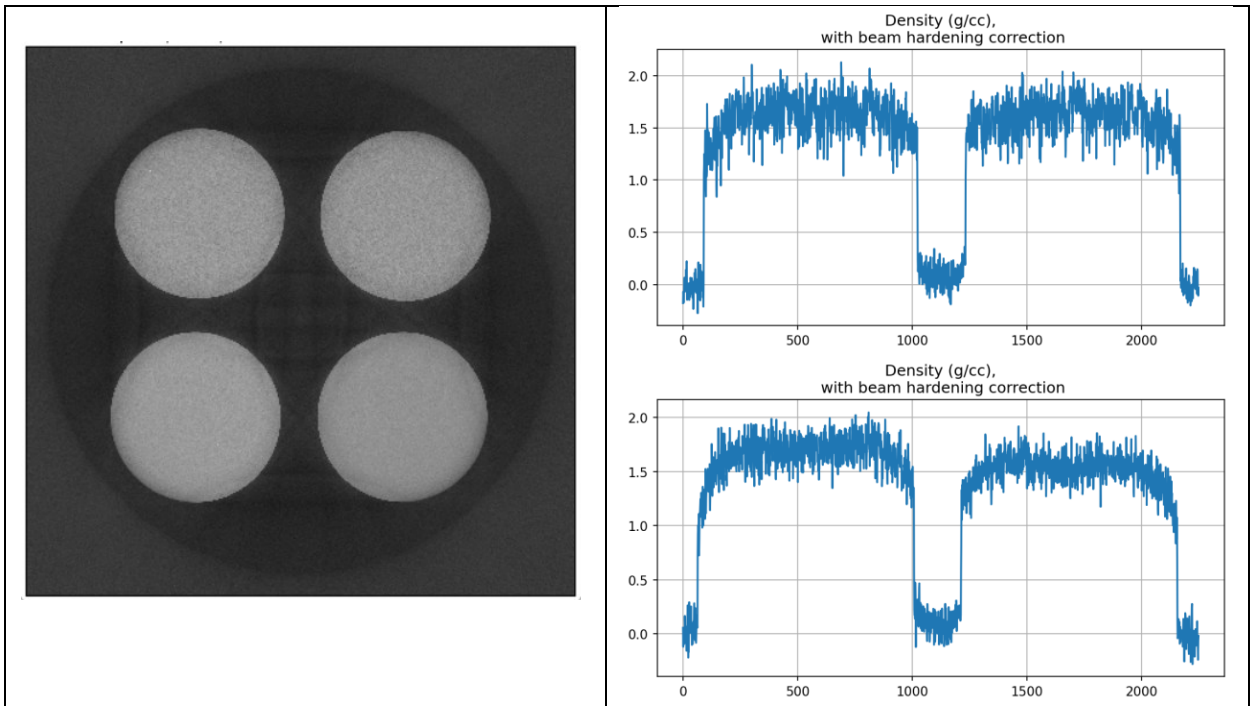

Fig 6: Left: Slice from CT reconstruction of the graphite rods with beam hardening correction. Right top: Line plots through the top two cylinders, g/cm$^3$. Right bottom: Line plots through the bottom two cylinders, g/cm$^3$.

Graphite has a lot of internal structure, so the reconstructions have a lot of internal variation which appears as noise, but the average density as measured across the cross section of the cylinders is found to be in good agreement with manufacturer-reported densities. Table 1 shows the average density for the four graphite rods in a 2-dimensional slice from the reconstruction as shown in Fig 6.

| | |
|---|---|
| 1.66 g/cm³ (1.72 g/cm³) | 1.63 g/cm³ (1.78 g/cm³) |
| 1.70 g/cm³ (1.86 g/cm³) | 1.56 g/cm³ (1.66 g/cm³) |

Table 1: Estimated densities of the four cylinders made of four commercially available graphite materials. Text in parentheses are the densities as reported by the manufacturer. The underestimation ranges from 3.4% to 8.6%.

Some possible sources of error are:

- Underestimation of the density due to the presence of scatter.
- Differences in geometry between SR measurements and OUT scans.
- Higher ash content in some graphite samples than others such that they are not strictly made of the same material as the SR.

## Summary

We have shown that in the case of single material objects where well-characterized reference material sheets are available, one can obtain beam hardening-corrected reconstructions and absolute density numbers with good accuracy (<10% error). This can be lowered if the OUT and SR have precisely the same chemical composition. Another known source of error is scatter, which causes an underestimation of the density. Future work to improve accuracy by reducing scatter and other sources of error is being considered.

## Acknowledgements

The author thanks Dr. Adam Denny for providing excellent experimental data. The author acknowledges Dr. Aaron Luttman and Dr. Luke Campbell for helpful discussions and insightful feedback.